\documentclass[reprint, aps, pra, superscriptaddress]{revtex4-2}

\usepackage{hyperref}
\hypersetup{
	colorlinks=true,
	linkcolor=blue,
	filecolor=blue,      
    urlcolor=blue,
}
\usepackage{amsmath,amssymb,graphicx}
\usepackage{siunitx,bm}
\usepackage{booktabs}

\begin{document}

\title{Characterization of the electronic ground state of Mg$_2^+$ by PFI-ZEKE photoelectron spectroscopy}

\author{Matthieu Génévriez}
\affiliation{Institute of Condensed Matter and Nanosciences, Université catholique de Louvain, Louvain-la-Neuve B-1348, Belgium}
\affiliation{Laboratory of Physical Chemistry, ETH Zurich, CH-8093 Zurich, Switzerland}
\author{Maxime Holdener}
\author{Carla Kreis}
\author{Frédéric Merkt}
\affiliation{Laboratory of Physical Chemistry, ETH Zurich, CH-8093 Zurich, Switzerland}

\begin{abstract}
The $\mathrm{X}^+\ ^2\Sigma_u^+$ ground electronic state of the Mg$_2^+$ ion
has been studied by pulsed-field-ionization zero-kinetic-energy (PFI-ZEKE)
photoelectron spectroscopy. Rotationally cold Mg$_2$ molecules produced in a
laser-ablation supersonic-expansion source were ionized by resonant two-photon
absorption from the $\mathrm{X}\ ^1\Sigma_g^+(v''=0)$ ground vibronic state.
The photoelectron spectra of the $v^+=3-14$ vibrational levels of the
$\mathrm{X}^+$ state were recorded with rotational resolution and their
analysis led to the determination of accurate term values and rotational
constants for these levels. Improved values of the adiabatic ionization energy
of $^{24}$Mg$_2$ ($51\,503.9(4)$~cm$^{-1}$) and of the ground-state
dissociation energy of $^{24}$Mg$_2^+$ ($10 572.3(6)$~cm$^{-1}$) were
determined from experimental data.
\end{abstract}

\maketitle

\section{Introduction}

Alkaline-earth-metal diatomic molecules have been extensively studied (see,
\textit{e.g.}, \cite{allard02,knockel13,heaven11,stein08,li13}). They have a
very weakly bound ground electronic state and unusual chemical properties~\cite
{heaven11}. For example, simple molecular-orbital theory fails to describe the
chemical bond of the ground state of Be$_2$~\cite{roeggen96,bondybey84} and the
calculation of the potential-energy functions of this few-electron molecule to
the level of accuracy reached experimentally is still a challenge for
\textit{ab initio} quantum chemical methods~\cite{merritt09,heaven11}. The
alkaline-earth-metal atoms (M$=$Be, Mg, Ca, Sr, Ba) and their singly charged
ions M$^+$ can easily be laser
cooled~\cite{kurosu92,beverini89,xu03,neuhauser78,wineland78} and the accurate
knowledge of the properties of the alkaline-earth-metal dimers (M$_2$) and
their ions (M$_2^+$) is important for characterizing atomic collisions in
ultracold gases. For example, scattering lengths are related to the properties
of the molecular potential-energy functions at large internuclear distances
(see Ref.~\citenum{tiesinga02} for an example with Mg$_2$). The accurate
knowledge of molecular structure is also required when designing
photoassociation schemes used to produce molecules in ultracold atomic
ensembles~\cite {nagel05,zinner00,jones06}. For these reasons, the ground and
first-excited states of Mg$_2$ have been thoroughly characterized by
high-resolution spectroscopy, and highly accurate potential-energy functions
were derived from experimental data~\cite{knockel13,knockel14}, building upon
earlier measurements of the absorption and laser-induced-fluorescence spectra
of Mg$_2$~\cite{scheingraber77,balfour70a,knight76,lauterwald92}.

In contrast to the neutral species, little is known on the singly charged
alkaline-earth-metal diatomic cations M$_2^+$. The ground electronic state of
Be$_2^+$ was investigated by pulsed-field-ionization zero-kinetic-energy
photoelectron spectroscopy (PFI-ZEKE-PES)~\cite{antonov10} and the ground-state
potential-well depth ($D_\mathrm{e}$) of Sr$_2^+$ was determined from the
photoionization spectra of Sr$_2$~\cite {dugourd92}. Experimental
investigations of Mg$_2^+$ are limited to a photodissociation study, which led
to the determination of the ground-state dissociation energy ($D_0 = 10\,200
\pm 300$~cm$^ {-1}$)~\cite{ding93}. Most of what is known on this ion was
obtained in high-level \textit{ab initio} calculations~\cite{li13,sodupe92,ricca94,ladjimi18,alharzali18,smialkowski20a,smialkowski20b}.
The development of hybrid ion-atom traps, accommodating both ultracold atomic
gases and trapped ions~(see, \textit{e.g.}, Refs.~\citenum{smith05a,ratschbacher12,eberle16}) has prompted the need of further
studies of alkaline-earth-metal dimer ions because, as for the neutral molecule
case, molecular potential-energy functions play an important role in the
description of cold collisions and photoassociation within the
trap~\cite{tomza19}.

The knowledge of the ground electronic state of Mg$_2^+$ is also required as a
starting point to study the electronically excited states of the ion, and in
particular the Rydberg series converging to the metastable levels of the
Mg$_2^{2+}$ ion~\cite{hogreve04}, which have not been observed so far. We plan
to study these states following the procedure we used recently to characterize
the ground electronic state of MgAr$^{2+}$ and the Rydberg series of MgAr$^+$
converging to this doubly charged
ion~\cite{genevriez20,wehrli20a,genevriez20a,wehrli21b,genevriez21c,wehrli21}.

We report an experimental study of the $\mathrm{X}^+\ ^2\Sigma_u^+$ ground
electronic state of Mg$_2^+$ by high-resolution
PFI-ZEKE-PES~\cite{reiser88,muller-dethlefs91,chupka93,merkt11}. Ground-state
Mg$_2$ molecules formed in a laser-ablation supersonic-expansion source were
excited to the region of the Mg$_2^+$($\mathrm{X}^+$) ionization threshold by
resonant two-photon excitation, as described in detail in
Sec.~\ref{sec:experiment}. Rotationally resolved photoelectron spectra were
recorded for the $v^+=3-14$ vibrational levels of the ion. They are
presented and analyzed in Sec.~\ref{sec:results}. Molecular constants
including the dissociation energy of Mg$_2^+(\mathrm{X}^+)$, its vibrational
and rotational constants and the adiabatic ionization potential of Mg$_2$ are
derived and compared to available data. Our results provide a comprehensive and
accurate description for the ground-electronic state of Mg$_2^+$.

\section{Experiment}\label{sec:experiment}

The experimental setup used in the present work is based on the one used to
study Mg and MgAr ions, as described in Refs.~\citenum {genevriez19,wehrli20},
and was adapted to the production of Mg$_2$ molecules and ions. Mg$_2$ dimers
are formed by laser ablation of a Mg rod within the nozzle of a
supersonic-expansion source with Ne used as carrier gas. Ablation is carried
out by focusing the second harmonic of a nanosecond pulsed Nd:YAG laser
(repetition rate 25~Hz) onto the Mg rod with an $f=30$~cm lens. A pulse energy
of $\sim 10$~mJ was found to be an optimal compromise to simultaneously reach a
large Mg$_2$ number density, low shot-to-shot fluctuations and a long-term
stability of the ablation process. As is common for metal-cluster
laser-ablation sources~\cite{duncan12}, the nozzle was chosen to be a narrow
(1-mm diameter) and long (15-mm) tube along the molecular-beam-propagation direction such
that the metal-vapor density remained large over a significant distance to
increase metal-dimer formation. We observed that this growth channel
significantly reduces the shot-to-shot fluctuations of the Mg$_2$ density in
the molecular beam. 6.5~cm beyond the nozzle orifice, the molecular beam is
collimated by a 3-mm-diameter skimmer before it enters the photoexcitation
chamber.

\begin{figure}
	\centering
	\includegraphics[width=0.95\columnwidth]{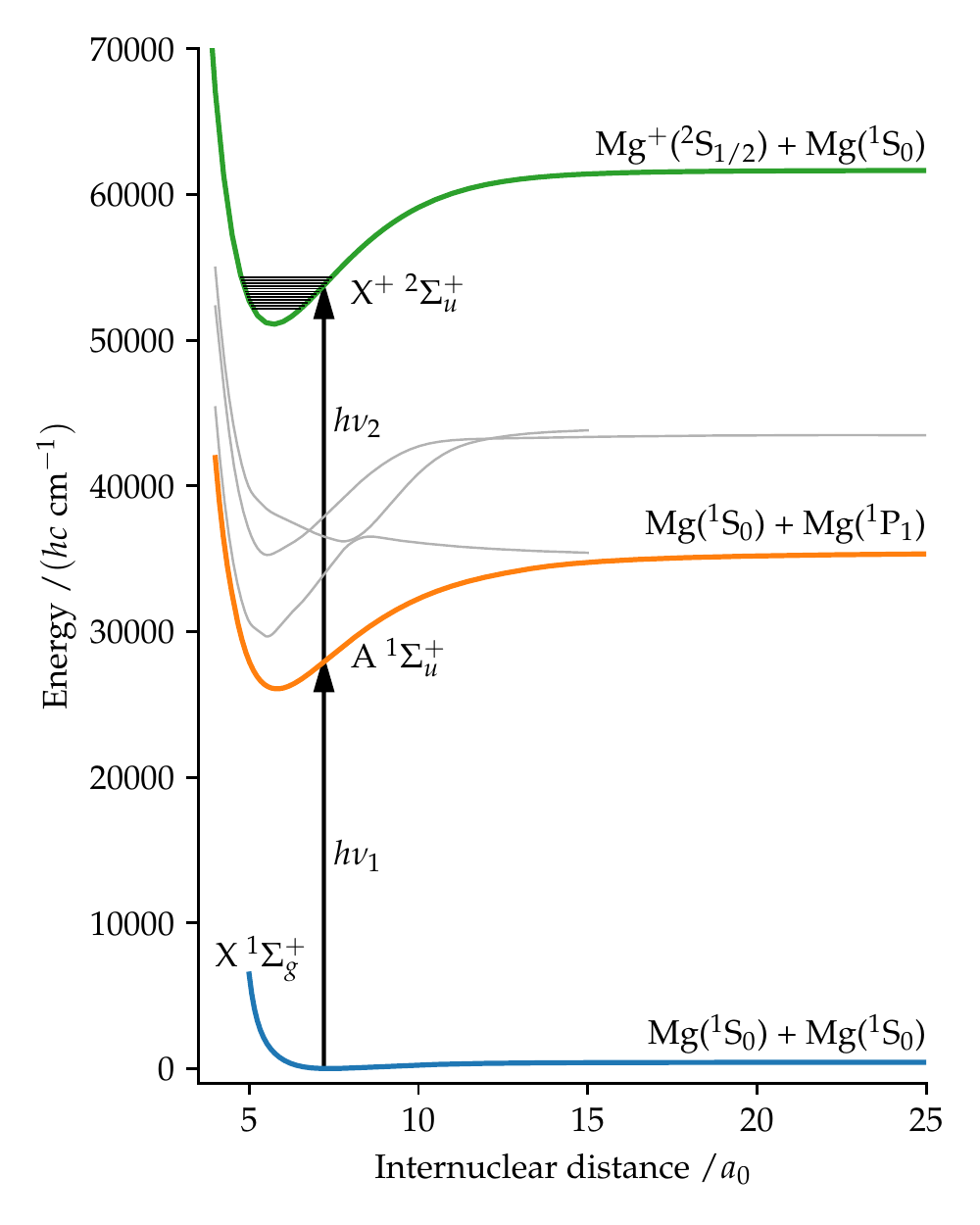}
	\caption{Potential-energy functions of the electronic states of Mg$_2$ and of the ground electronic state of Mg$_2^+$ relevant for the present investigation. The functions of the X and A states of Mg$_2$ were calculated using the model potentials and parameters from Refs.~\citenum{knockel13} and~\citenum{knockel14}, respectively. The gray potential-energy curves correspond to higher-lying singlet ungerade states of Mg$_2$ calculated in Ref.~\citenum{amaran13}. The potential-energy function of the X$^+$ state of Mg$_2^+$ is from Ref.~\citenum{smialkowski20a}. The two-photon ionization scheme used in the experiment is illustrated by the vertical arrows. The full horizontal lines in the X$^+$-state potential-energy function represent the vibrational levels measured in the present work.}
	\label{fig:potential_energy_functions}
\end{figure}

In this chamber, the molecular beam is intersected at right angles by two
tuneable dye lasers pumped by the second harmonic of a seeded Q-switched Nd:YAG
laser. The photoexcitation scheme depicted in
Fig.~\ref{fig:potential_energy_functions} is used to excite the Mg$_2$
molecules from their $\mathrm{X}\ ^1\Sigma_g^+$ ground state to the region of
the first ionization thresholds using two tuneable lasers in a resonant
($1+1'$) two-photon absorption process. The first dye laser, called Laser 1
hereafter, is operated with Pyridine 1 and 2 dyes and its fundamental output is
frequency-doubled in a beta-barium-borate (BBO) crystal to generate radiation
with wavenumbers tuneable in the range from $\tilde{\nu}_1=27000$~cm$^{-1}$ to
$28000$~cm$^{-1}$. The frequency-doubled output is used to excite Mg$_2$
molecules in their electronic ground state ($\mathrm {X}\ ^1\Sigma_g^+$) to the
first electronically excited state ($\mathrm {A}\ ^1\Sigma_u^+$). The second
laser, called Laser 2, is operated with Styryl 11 dye and its output is also
frequency doubled in a BBO crystal to generate radiation in the range from
$\tilde{\nu}_2=24500$~cm$^{-1}$ to $26500$~cm$^ {-1}$. This radiation is used
to further excite the Mg$_2$ molecules from the $\mathrm{A}\ ^1\Sigma_u^+$
state to high Rydberg states located energetically just below rovibrational
levels of the $\mathrm{X}^+\ ^2\Sigma_u^+$ ground electronic state of the
Mg$_2^+$ molecular ion. The wavenumbers of the fundamental outputs of both dye
lasers were calibrated using a commercial wavemeter with a specified absolute
accuracy of 0.02~cm$^{-1}$. The pulse energy of the frequency-doubled output of
Laser 1 was kept below $\sim 50$~$\mu$J to avoid power broadening of the
$\mathrm{A} \leftarrow \mathrm{X}$ transition and ionization of molecules in
the A state by absorption of a further photon of the same wavenumber.

Photoexcitation occurs within a stack of 5 cylindrical resistively coupled
electrodes surrounded by two concentric mu-metal shields to suppress stray
magnetic fields. Electric potentials applied to the stack generate electric
fields in the interaction region that serve either to field-ionize molecules in
high Rydberg states and accelerate the electrons into a flight tube or to
extract the Mg$_2^+$ ions produced by photoionization, depending on the sign of
the potential difference. Charged particles are detected at the end of the
flight tube by a microchannel-plate detector. The different masses, and in
particular those associated with the various Mg$_2^+$ isotopomers ($^ {24, 25,
26}$Mg), are separated by their times of flight. Natural abundances $N_ {m_1,
m_2}$ of the $^{m_1}\mathrm{Mg}^{m_2}\mathrm{Mg}$ isotopomers can be calculated
from those of the Mg isotopes ($N_{24}=0.7899(4)$, $N_{25}=0.1000(1)$, and
$N_{26}=0.1101(3)$~\cite{rosman98}), which gives $N_{24, 24} = 0.624, N_{24,
25} = 0.158, N_{24, 26} = 0.174, N_{25, 25} = 0.010, N_{25, 26} = 0.022$ and
$N_{26, 26} = 0.012$.

To record spectra of the $\mathrm{A} \leftarrow \mathrm{X}$ transition of
Mg$_2$, the Mg$_2^+$ ions produced by ($1+1'$) resonance-enhanced multiphoton
ionization (REMPI) are extracted using a single large electric-field pulse of
$+240$~V\,cm$^{-1}$. The ion signals corresponding to the different Mg$_2^+$
isotopomers are monitored as a function of the wavenumber of the doubled output
of Laser 1. To record PFI-ZEKE-PE spectra, the wavenumber of Laser 1 is kept
fixed so as to select a specific rovibrational level of the A state. The
doubled wavenumber of Laser 2 is scanned across the $\mathrm{X}^+ \leftarrow
\mathrm{A}$ ionization thresholds while monitoring the yield of electrons
generated by delayed pulsed field ionization of very high Rydberg states
(principal quantum number $\gtrsim 200$) located just below the successive
ionization thresholds~\cite{muller-dethlefs98}. Sequences of small
electric-field pulses are applied to the electrode stack to field ionize the Rydberg
states and accelerate the electrons towards the detector. The number of pulses
in the sequence, their amplitudes, and their polarities determine the resolution
of the PFI-ZEKE-PE spectra and the signal strength~\cite{hollenstein01}. After
systematic optimization, we found that the sequences ($0.09, -0.09, -0.17,
-0.26, -0.35, -0.43, -0.52$) V\,cm$^ {-1}$ and ($0.17, -0.12, -0.21, -0.29,
-0.36$) V\,cm$^{-1}$ provided the best spectra for weak and strong bands of the
photoelectron spectrum, respectively, resulting in spectral resolutions of
$0.5$ and $0.6$~cm$^{-1}$, respectively.

The lines recorded in PFI-ZEKE-PE spectra correspond to high Rydberg states of
the neutral molecule located just below the ionization thresholds associated
with specific rovibronic levels of the molecular ion. To determine the position
of the field-free ionization thresholds, the spectra were corrected for these
field-induced shifts following the procedure described in Refs.~\citenum
{hollenstein01,wehrli21a}. We verified the accuracy of the correction procedure
by recording the PFI-ZEKE-PE spectrum of metastable Mg($\mathrm{3s3p}\ ^3P_
{1}$) atoms produced in the source. The ionization energy of this state is well
known (39800.59(4)~cm$^{-1}$~\cite{kramida20}), and we found that the
correction of the field-induced shifts is accurate to within 0.04~cm$^ {-1}$, a
value more than 10 times smaller than the full width at half maximum of the
lines of the photoelectron spectra ($\sim 0.5$~cm$^{-1}$).

\section{Results}\label{sec:results}

\subsection{($1+1'$) REMPI spectra of the $\mathrm{A} \leftarrow \mathrm{X}$ transition of Mg$_2$}

\begin{figure}
	\includegraphics[width=\columnwidth]{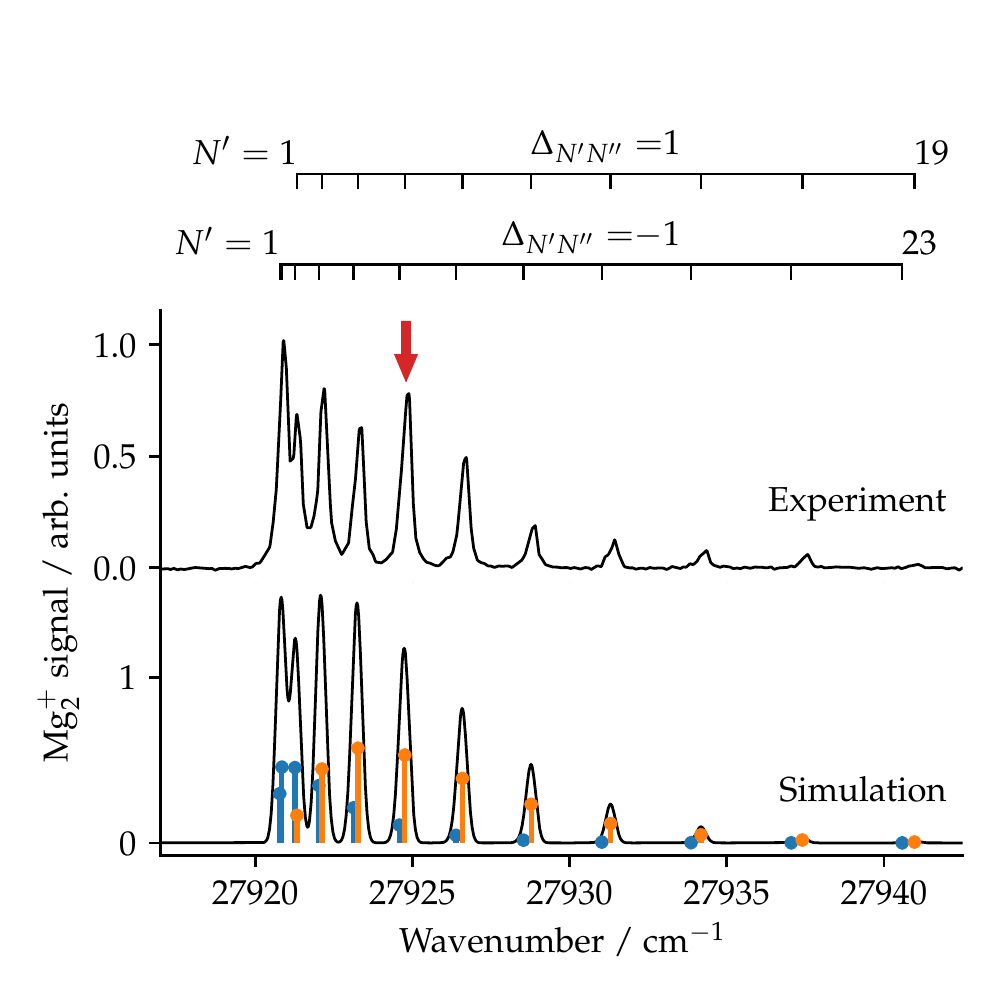}
	\caption{Measured (top) and calculated (bottom) REMPI spectra of the $\mathrm{A}\ ^1\Sigma_u^+(v'=10) \leftarrow \mathrm{X}\ ^1\Sigma_g^+ (v''=0)$ transition of $^{24}\mathrm{Mg}^{24}\mathrm{Mg}$. The assignment bars above the spectra give the positions of transitions to rotational levels of the $\mathrm{A}\ ^1\Sigma_u^+(v'=10)$ state for successive values of $N'$. The stick spectrum in the lower panel shows the wavenumbers and relative intensities of individual rovibronic transitions for a rotational temperature of 7~K. Individual lines in the calculated spectrum have a Gaussian shape with a full width at half maximum of 0.35~cm$^{-1}$. The vertical arrow indicates the wavenumber at which the first laser was set when recording the PFI-ZEKE-PE spectra presented in Fig.~\ref{fig:zeke_spectrum}.}
	\label{fig:rempi_spectrum}
\end{figure}

The ground-state Mg$_2$ molecules produced in the experiment were characterized
by recording ($1+1'$) REMPI spectra of the $\mathrm{A}\leftarrow\mathrm{X}$
transition. A typical spectrum is shown in Fig.~\ref{fig:rempi_spectrum} and
corresponds to the $\mathrm{A}\ ^1\Sigma_u^+(v'=10) \leftarrow \mathrm
{X}\ ^1\Sigma_g^+ (v''=0)$ vibrational band.

\begin{figure*}
	\includegraphics[width=1.3\columnwidth]{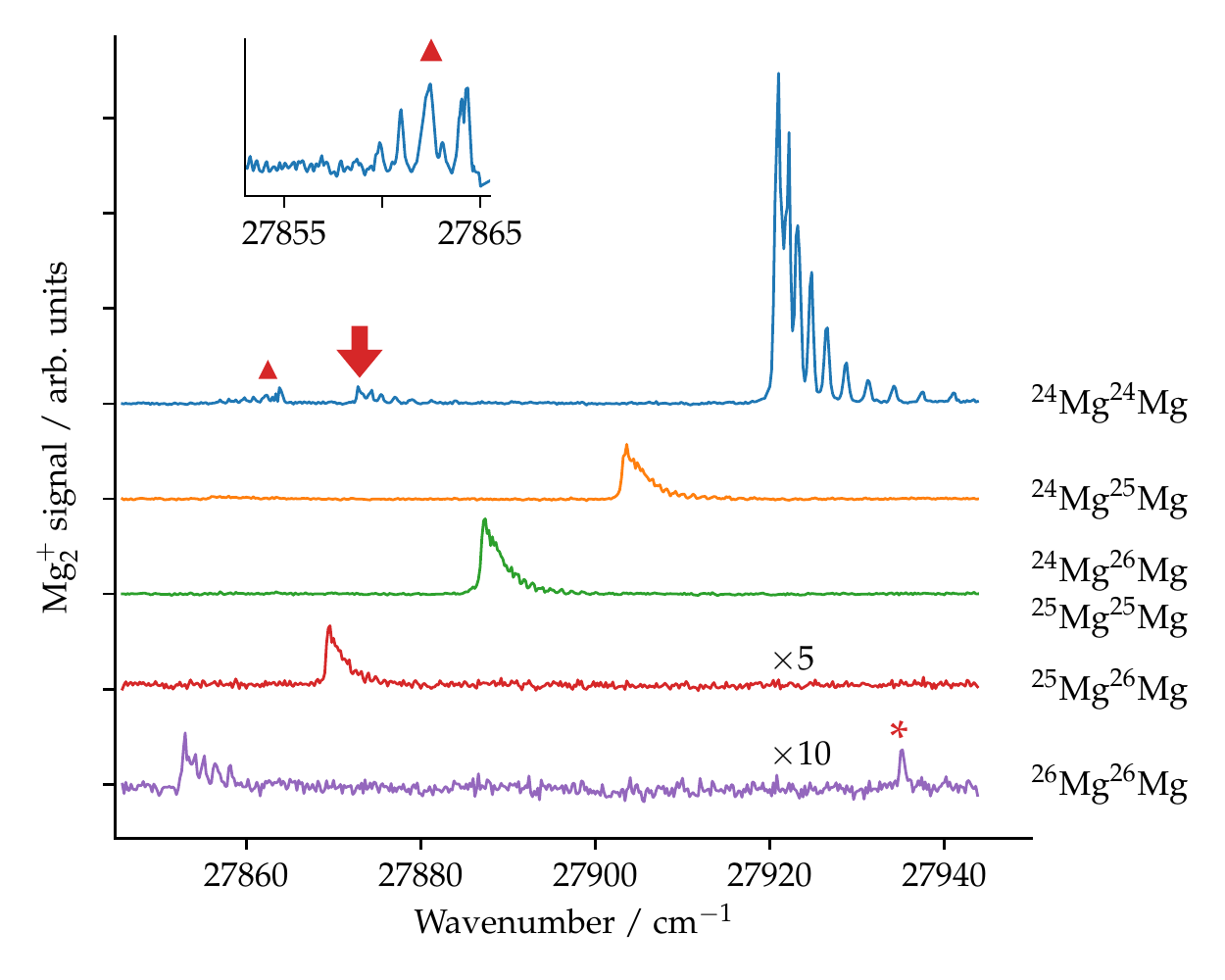}
	\caption{Spectra of the $\mathrm{A}\ ^1\Sigma_u^+(v'=10) \leftarrow \mathrm{X}\ ^1\Sigma_g^+ (v''=0)$ transition of the different isotopomers of $\mathrm{Mg}_2$, as labeled on the right of the figure. The vertical arrow shows the position of the $\mathrm{A}\ ^1\Sigma_u^+(v'=10) \leftarrow \mathrm{X}\ ^1\Sigma_g^+ (v''=1)$ hot band for $^{24}\mathrm{Mg}_2$. The line labeled with an asterisk could not be assigned. The band marked by a triangle in the $^{24}\text{Mg}_2$ spectrum is tentatively assigned to excitation from the $(1)\ ^3\Sigma_u^+$ state to a high-lying excited state (see text). A spectrum of the same band recorded at higher resolution is shown in the inset on the top left of the figure. The vertical scales of the spectra of $^{25}\mathrm{Mg}^{26}\mathrm{Mg}$ and $^{26}\mathrm{Mg}_2$ are magnified by a factor 5 and 10, respectively, for visibility. The spectra have been shifted along the vertical axis for clarity.}
	\label{fig:rempi_spectrum_isotopes}
\end{figure*}

The $\mathrm{X}\ ^1\Sigma_g^+$ and $\mathrm{A}\ ^1\Sigma_u^+$ states are well
described by Hund's angular-momentum-coupling case (b), with $N''$ and $N'$
representing the total-angular-momentum-without-spin quantum numbers of the
initial and final states, respectively. Each vibrational band of this $\Sigma -
\Sigma$ transition thus consists of P ($\Delta N = N' - N'' = -1$) and R
($\Delta N= 1$) rotational branches. Because of nuclear-spin statistics, the
homonuclear Mg$_2$ isotopomers with zero nuclear spin ($^{24}\mathrm{Mg}_2$ and
$^{26}\mathrm{Mg}_2$) can only occupy rotational levels of even (odd) values of
$N$ for gerade (ungerade) electronic states and their spectra thus exhibit
twice fewer lines than the other isotopomers.
Figure~\ref{fig:rempi_spectrum_isotopes} shows overview spectra of the $\mathrm
{A}\ ^1\Sigma_u^+(v'=10) \leftarrow \mathrm{X}\ ^1\Sigma_g^+(v''=0)$ transition
for all Mg$_2$ isotopomers. The spectra were recorded at high laser pulse
energy and the lines are slightly power broadened. The rotational structure of
the vibrational bands is resolved for the $^ {24}\mathrm{Mg}_2$ and
$^{26}\mathrm{Mg}_2$ species only because of the smaller number of rotational
transitions resulting from nuclear-spin statistics. The rotational structure of
the band of the homonuclear $^ {25}\mathrm{Mg}_2$ molecule
($I_{^{25}\mathrm{Mg}} = 5/2$) cannot be unambiguously observed in the spectra
because it possesses the same mass and approximately the same isotopic shift as
the $^{24}\mathrm{Mg}^{26}\mathrm{Mg}$ isotopomer while its natural abundance
is more than $10$ times smaller. When taking laser-pulse-energy variations
across the spectral range into account, the amplitudes of the spectra match the
relative natural abundances of the various isotopomers.

The REMPI spectra were modelled using band origins and rotational constants
calculated with the accurate potential-energy functions of the $\mathrm {X}\
^1\Sigma_g^+$ and $\mathrm{A}\ ^1\Sigma_u^+$ electronic states determined from
high-resolution spectroscopic data by Kn\"ockel \textit{et al.}~\cite
{knockel13,knockel14}. At the low rotational temperature of the supersonic
expansion, only rotational levels with $N'' \leq 24$ are occupied and
centrifugal distortion effects are negligible within the experimental accuracy.
The agreement between the observed and calculated line positions is excellent.
The line intensities of the individual $\mathrm {A}\ ^1\Sigma_u^+(v', N')
\leftarrow \mathrm{X}\ ^1\Sigma_g^+ (v'', N'')$ transitions were calculated
from the line strengths for transitions between states described by Hund's
angular-momentum-coupling case (b)~\cite {zare88}. Rotational levels of the
X($v=0$) initial state were assumed to be thermally populated, and the
theoretical spectrum matched the experimental one best for a rotational
temperature of $\sim 7$~K. The vibrational temperature of the ground-state
molecules in the beam was derived from the relative amplitudes of the
$\mathrm{A}\ ^1\Sigma_u^+(v') \leftarrow \mathrm{X}\ ^1\Sigma_g^+ (0)$ band and
the $\mathrm{A}\ ^1\Sigma_u^+(v') \leftarrow \mathrm {X}\ ^1\Sigma_g^+ (1)$ hot
band (see Fig.~\ref {fig:rempi_spectrum_isotopes} for the $^{24}\mathrm{Mg}_2$
isotopomer) taking into account the associated Franck-Condon factors calculated
using the potential-energy functions from Refs.~\citenum{knockel13,knockel14}. The
resulting value of $\sim 30$~K indicates that the cooling of the vibrational
degree of freedom in the supersonic expansion is less efficient than the
cooling of the rotational ones.

The band marked by a triangle in Fig.~\ref{fig:rempi_spectrum_isotopes} does
not belong to the $\mathrm{A}\ ^1\Sigma_u^+(v') \leftarrow \mathrm {X}\
^1\Sigma_g^+ (v'')$ band system. Unlike the other bands we recorded, it
degrades to lower wavenumbers, rather than to higher wavenumbers. Consequently,
the rotational constant of the upper state is smaller, and its mean
internuclear distance larger, than in the initial state. In the wavenumber
range considered, no other electronic state can be reached by dipole excitation
from the $\mathrm{X}\ ^1\Sigma_g^+ (v'' \le 4)$ levels. In particular, the
$(1)\ ^1\Pi_u$ state investigated in Refs.~\citenum{lauterwald92,knockel14} is not energetically accessible. Transitions from the ground state to triplet states
cannot give rise to the observed band either, because only the dissociation
continua of the low-lying triplet states of Mg$_2$ correlating to the
$\text{Mg}(3s^2\ ^1S) + \text{Mg}(3s3p\ ^3P)$ dissociation asymptote are
accessible at the photon energies used to record the spectra.

Metastable molecules are known to form in laser-ablation supersonic-expansion
sources, an example being the production of MgAr molecules in their
$\mathrm{a}\ ^3\Pi_0$ state~\cite{bennett90}. Mg$_2$ molecules in metastable
states may be formed in our experiment, in particular the low-lying $(1)\
^3\Pi_g$ and $(1)\ ^3\Sigma_u^+$ states correlating to the $\text{Mg}(3s^2\
^1S) + \text{Mg} (3s3p\ ^3P)$ dissociation asymptote and the $(1)\ ^1\Pi_g$
state correlating to the $\text{Mg}(3s^2\ ^1S) + \text{Mg} (3s3p\ ^1P)$
dissociation asymptote (see, \textit{e.g.}, Ref.~\citenum{czuchaj01}).
Fluorescence of the $^1\Pi_g$ and $^3\Pi_g$ states to the $\mathrm {X}\
^1\Sigma_g^+$ ground state is forbidden because of the $g \leftrightarrow u$
selection rule whereas fluorescence of the $^3\Sigma_u^+$ state is forbidden
by the approximate $\Delta S = 0$ selection rule. The $(1)\ ^3\Sigma_u^+$ state
can in principle radiate to the ground state \textit{via} spin-orbit mixing
with $^1\Pi_u$ states, but its radiative lifetime is not known. \textit{Ab
initio} calculations predict that these three states are the lowest-lying
states of their respective electronic symmetries and that they are strongly
bound~\cite{czuchaj01}.

Because of nuclear-spin symmetry, only odd values of $N''$ are allowed for the
$^3\Sigma_u^+$ state whereas both even and odd values of $N''$ can be populated
in the case of the $^{1,3}\Pi_g$ states. We have observed in
Fig.~\ref{fig:rempi_spectrum_isotopes} that the existence of both even and odd
$N''$ values in the initial state leads to unresolved rotational structures of
the bands whereas the rotational structure is partially resolved when only even
(odd) values are allowed (compare, for example, the spectra of the
$^{24}\text{Mg}^{25}\text{Mg}$ and $^{24}\text{Mg}_2$ isotopomers). Because the
rotational structure of the band marked by a triangle in
Fig.~\ref{fig:rempi_spectrum_isotopes} is partially resolved, we tentatively
attribute it to excitation of metastable $^{24}\mathrm{Mg}_2$ molecules from
the $(1)\ ^3\Sigma_u^+$ state to a $^3\Sigma^+_g$ or $^3\Pi_g$ state in the
dense manifold of excited states lying $\sim 12\,000$~cm$^{-1}$ below the
Mg$_2^+$ $\mathrm{X}^+$ ionization threshold. This tentative assignment is
further supported by the fact that similar transitions were observed in Be$_2$
molecules produced in a laser-ablation supersonic-expansion
source~\cite{merritt08}. Specifically, transitions from the $(1)\ ^3\Sigma_u^+$
state of Be$_2$ to the $(2)\ ^3\Pi_g$ and $ (3)\ ^3\Pi_g$ state were observed
by REMPI spectroscopy in the same photon-energy range as the $\mathrm{B}\
^1\Sigma_u^+
\leftarrow \mathrm{X}\ ^1\Sigma_g^+$ and $\mathrm {A}\ ^1\Pi_u \leftarrow
\mathrm{X}\ ^1\Sigma_g^+$ transitions of Be$_2$~\cite{merritt08}. Further
investigations are required to clarify the assignment of this weak band.

\subsection{PFI-ZEKE-PE spectra of the $\mathrm{X^+} \leftarrow \mathrm{A}$ photoionizing transition of Mg$_2$}

\begin{figure*}
	\includegraphics[width=1.7\columnwidth]{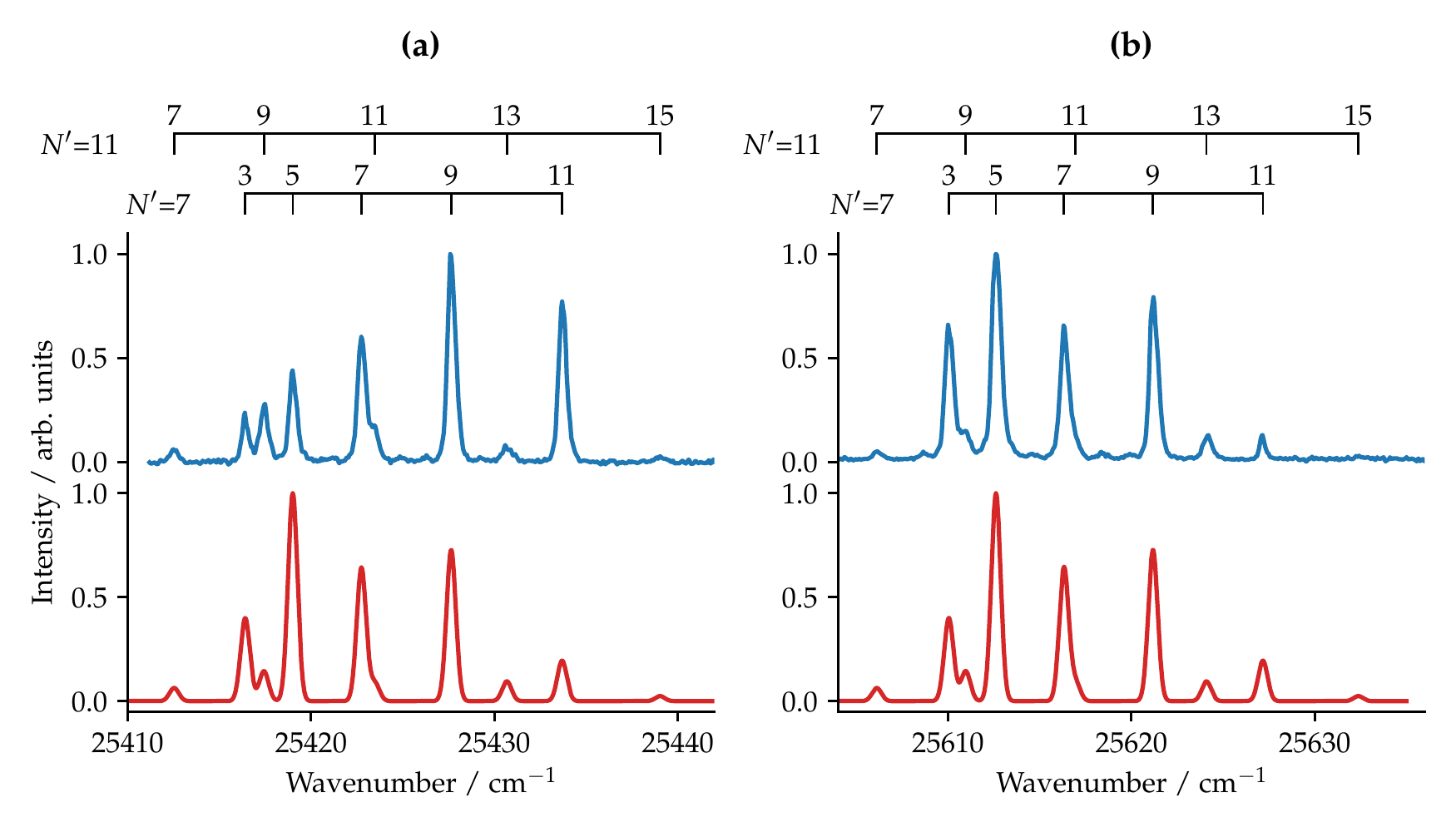}
	\caption{Measured (top) and calculated (bottom) PFI-ZEKE photoelectron spectra of (a) the $\mathrm{X}^+\ ^2\Sigma_u^+ (v^+=9) \leftarrow \mathrm{A}\ ^1\Sigma_u^+(v'=10)$ and (b) the $\mathrm{X}^+\ ^2\Sigma_u^+ (v^+=10) \leftarrow \mathrm{A}\ ^1\Sigma_u^+(v'=10)$ photoionizing transitions of $^{24}\text{Mg}_2$. The assignment bars indicate transitions from initial rotational states with $N'=7$ and $11$ to states of the ion with successive values of $N^+$. The dominant branches correspond to $N^+ - N' = 0, \pm 2, \pm 4$. Individual lines in the calculated spectra have a Gaussian shape with a full width at half maximum of 0.6~cm$^{-1}$.}
	\label{fig:zeke_spectrum}
\end{figure*}

PFI-ZEKE-PE spectra of the $\mathrm{X}^+\ ^2\Sigma_u^+ (v^+, N^+) \leftarrow
\mathrm{A}\ ^1\Sigma_u^+(v', N')$ photoionizing transitions of the
$^{24}\mathrm{Mg}_2$ isotopomer were recorded for $v^+$ in the range from $3$
to $14$. For each value of $v^+$, $v'$ was chosen so that the Franck-Condon
factor of the transition was sufficiently large to record spectra with good
signal-to-noise ratios. The spectra of the $v^+=9 \leftarrow v'=10$ and $v^+=10
\leftarrow v'=10$ transitions are presented as examples in Fig.~\ref
{fig:zeke_spectrum}.

In the experiment, Mg$_2$ in its ground electronic and vibrational state was
first photoexcited to selected $\mathrm {A}\ ^1\Sigma_u^+(v', N')$
rovibrational states. An example is illustrated by the red arrow in
Fig.~\ref{fig:rempi_spectrum}, which indicates the wavenumber at which the
first laser ($\tilde{\nu}_1 = 27924.8$~cm$^{-1}$) excites ground-state
molecules with $v''=0$ and $N''=6, 12$ to levels of the A state with $v'=10$
and $N'=7$ and 11, respectively. PFI-ZEKE-PE spectra were then recorded from
these two levels by scanning the wavenumber of Laser 2 in a range just below
the $\mathrm{X}^+\ ^2\Sigma_u^+ (v^+=9 - 10, N^+)$ ionization thresholds. The
spectra associated with each electric-field pulse of the PFI sequence were
corrected for the field-induced shifts, as described in
Sec.~\ref{sec:experiment}, and summed to yield the spectra depicted in the top
panels of Fig.~\ref{fig:zeke_spectrum}.

These spectra exhibit more rotational branches than the P
($\Delta N = -1$) and R ($\Delta N = + 1$) branches characteristic of $\Sigma -
\Sigma$ transitions between bound states. This observation originates from the
fact that the photoelectron also carries angular momentum, which needs to be
considered when deriving photoionization selection
rules~\cite{xie90,watson98,signorell97}. The photoelectron is described as a
superposition of partial waves with angular-momentum quantum numbers $l$. The
parity of the final-state wavefunction is determined by the $g/u$ symmetry of
the ion core, which is ungerade in the present case, the parity of the Rydberg
electron, which is given by $ (-1)^l$, and the parity of the ion-core
rotational function, which is $(-1)^{N^+}$. The transitions must obey the
rovibronic parity selection rule~\cite{xie90,watson98,signorell97}
\begin{align}
	\pm \leftrightarrow \pm &\qquad \text{for}\ l\ \text{odd}, \\
	\pm \leftrightarrow \mp &\qquad \text{for}\ l\ \text{even},
\end{align}
where the $\pm$ signs give the rovibronic parity of the neutral and ionized molecules. Because both the A state of Mg$_2$ and the X$^+$ state of Mg$_2^+$ have $u$ electronic symmetry, the rotational selection rule can be expressed as
\begin{align}
	\Delta N = N^+ - N' = 0, \pm 2, \pm 4, \ldots &\qquad \text{for}\ l\ \text{odd}, \\
	\Delta N = \pm 1, \pm 3, \pm 5, \ldots &\qquad \text{for}\ l\ \text{even}.
\end{align}
Upon photoionization, an electron is removed from the $3p \sigma_g$ orbital of
the A state with electronic configuration
$(3s\sigma_g)^2(3s\sigma_u^*)^1(3p\sigma_g)^1$. Consequently, the photoelectron
partial waves must have an odd $l$ value, and $\Delta N$ must be even.

For the homonuclear isotopomers, the parity selection rule gives the same
result as the nuclear-spin-symmetry conservation rule. Rotational levels of
$^{24}\text{Mg}_2$ and $^{26}\text{Mg}_2$
($I_{^{24}\text{Mg}}=I_{^{26}\text{Mg}}=0$) of ortho nuclear-spin symmetry
correspond to odd-$N$ levels in the A and X$^+$ states, so that $\Delta N$ must
be even. Rotational levels of $^{25}\text{Mg}_2$ ($I_{^{25}\text{Mg}}=5/2$) of
ortho (para) nuclear-spin symmetry have statistical weights 21 (15) and
correspond to even-$N$ (odd-$N$) levels in the A and X$^+$ states. For the
heteronuclear isotopomers, there are no restrictions from the conservation of
nuclear-spin symmetry, and the $g/u$ symmetry is only approximate. We have,
however, observed no evidence of such a symmetry breaking in the experimental
spectra we recorded.

The most intense lines in the experimental spectra are well described by 5
rotational branches corresponding to $N^+ - N' = \pm 0, \pm 2, \pm 4$, as shown
by the assignment bars in Fig.~\ref{fig:zeke_spectrum}. The line positions
were determined from the experimental spectra from fits using Gaussian
functions. The band origins $\tilde{\nu}_{v^+v'}$ and the rotational
constants $B^+_{v^+}$ of the final states were then obtained from the line
positions in a least-squares fit based on the standard
expression~\cite{herzberg50}

\begin{equation}
	\tilde{\nu} = \tilde{\nu}_{v^+v'} + B^+_{v^+} N^+\left(N^++1\right) - B'_{v'} N' \left(N' +1\right) .
	\label{eq:lineposition}
\end{equation}
The rotational constant $B'_{v'}$ of the A state was fixed to the value
obtained from the potential-energy function of Ref.~\citenum{knockel14}. The
values obtained from the fit are listed in Table~\ref{tab:lines}.

The intensities of the various
$\mathrm{X}^+\ ^2\Sigma_u^+(v^+, N^+) \leftarrow \mathrm {A}\ ^1\Sigma_u^+(v',
N')$ ionizing transitions can be estimated using a model developed by
Buckingham, Orr and Sichel~\cite{buckingham97} (see also
Ref.~\citenum{merkt93a}). The model relies on a single-center expansion of the
molecular orbital describing the bound electron before photoionization, from
which the following expression for the photoionization cross section can be
derived:

\begin{widetext}
	\begin{equation}
	\sigma_{N^+ \leftarrow N'} \propto  \sum_{l'=\lambda'}^\infty \frac{2N^++1}{2l'+1} 
	\begin{pmatrix}
		N^+ & l' & N' \\
		-\Lambda^+ & \lambda' & \Lambda'
	\end{pmatrix}^2 \left| c_{l'} \right|^2 \left[ l' \left| F_{\alpha, l'}^{E, l' -1} \right|^2 + (l'+1) \left| F_{\alpha, l'}^{E, l' +1} \right|^2 \right] .
	\label{eq:photoionization_xsec}
	\end{equation}
\end{widetext}
In Eq.~\eqref{eq:photoionization_xsec}, the quantum number $\lambda'$ is the
projection on the molecular axis of the orbital angular momentum $\bm{l'}$
associated with the single-center expansion of the bound molecular orbital. The
coefficients of the single-center expansion are represented by $\left| c_{l'}
\right|^2$ and the quantities $F_{\alpha, l'}^ {E, l}$ are bound-free radial
transition integrals. Here, we assume that, as in the hydrogen atom, $\left|
F_{\alpha, l'}^{E, l' +1} \right|^2 \gg \left| F_ {\alpha, l'}^{E, l' -1}
\right|^2$, such that the first term of the expression enclosed between the
square brackets on the right hand side of Eq.~\eqref {eq:photoionization_xsec}
can be neglected. The dependence of $\left| F_{\alpha, l'}^{E, l' +1}
\right|^2$ on the energy is assumed to be negligible within the range probed in
the experiment.

The lower panels of Fig.~\ref{fig:zeke_spectrum} show the spectra of the
$\mathrm{X}^+\ ^2\Sigma_u^+(9) \leftarrow \mathrm {A}\ ^1\Sigma_u^+(10)$ (left)
and $\mathrm{X}^+\ ^2\Sigma_u^+(10) \leftarrow \mathrm {A}\ ^1\Sigma_u^+(10)$
(right) transitions calculated using the results of a fit based on
Eq.~\eqref{eq:lineposition} and the expression for the line intensities given
by Eq.~\eqref {eq:photoionization_xsec}. In the latter equation, the
single-center expansion was truncated to $l' \le 4$ ($l'$ even), which yields
the 5 rotational branches observed in the experimental spectrum. The
coefficients $\left| c_{l'} F_{\alpha, l'}^{E, l' +1} \right|^2 = \left|
b_{l'}\right|^2$ were manually adjusted to reproduce as well as possible the
line intensities of the entire set of PFI-ZEKE-PE spectra we recorded, yielding
$\left| b_{0} \right|^2 = 0.09$, $\left| b_{2} \right|^2 = 0.57$ and $\left|
b_{4} \right|^2 = 0.34$. The populations of the initial rotational states
($N'=7, 11$) were determined from the calculated line intensities of the
respective $\mathrm{A}\ ^1\Sigma_u^+ (v', N') \leftarrow
\mathrm{X}\ ^1\Sigma_g^+ (v'', N'')$ transitions (vertical bars in the lower
panel of Fig.~\ref{fig:rempi_spectrum}).

The experimental line intensities are well reproduced by the calculation in the
case of the $\mathrm{X}^+\ ^2\Sigma_u^+(10) \leftarrow \mathrm {A}\
^1\Sigma_u^+(10)$ transition (right column in Fig.~\ref{fig:zeke_spectrum}), but the agreement is significantly worse for the $\mathrm{X}^+\ ^2\Sigma_u^+(9)
\leftarrow \mathrm {A}\ ^1\Sigma_u^+(10)$ transition (left column).
Equation~\eqref{eq:photoionization_xsec} assumes that direct photoionization is
predominant and disregards any interaction between the various ionization
channels. Such interactions couple ionization channels associated with
different ion-core rovibrational states and different photoelectron
orbital-angular momenta and energies, and give rise to phenomena such as
vibrational and rotational autoionization. Channel interactions can
significantly affect the rotational line intensities observed in PFI-ZEKE-PE
spectra~\cite{merkt93a}. Because the transition dipole moment scales as
$n^{-3/2}$ ($n$ is the principal quantum number), the coupling of Rydberg
states with different $n$ values modifies the photoexcitation probability.

In the spectra we measured, strong deviations of the line intensities from the
predictions of Eq.~\eqref{eq:photoionization_xsec} are observed for $v^+=3-9,
12, 14$. The intensities of the lines with positive $N^+-N'$ values are
systematically underestimated. When the effects of interactions between
different rotational channels, \textit{i.e.}, channels of different $N^+$
values, are significant, the intensities of the rotational branches with $N^+ -
N' > 0$ are expected to be reduced and those with $N^+-N' < 0$ to be
enhanced~\cite{merkt93a}, which is the opposite of what is observed in the
present case. Because the first electronically excited state of Mg$_2^+$ lies
more than 16\,000~cm$^{-1}$ above the ground state~\cite{ricca94}, electronic
channel interactions do not play a role. Consequently, we attribute the
observed deviations of the rotational line intensities from the model
predictions to predominantly vibrational channel interactions. Additional deviations can be caused by variations of the laser pulse energies or fluctuations of the density of Mg$_2$ molecules produced in the laser-ablation source. 

The values of the band origins $\tilde{\nu}_{v^+v'}$ and rotational constants
$B^+_{v^+}$ of the $\mathrm{X}^+(v^+)$ levels listed in Table~\ref{tab:lines}
were obtained from least-squares fits based on the line positions derived from
the spectra we measured. Spectra of the $v^+=13 \leftarrow v'=10$ vibrational band were recorded for different
initial rotational states ($N'=5, 7$) and the standard deviation of the band origins extracted from these was taken as the statistical uncertainties of all other band origins $\tilde{\nu}_{v^+v'}$ ($v^+ \neq 13$). 
The statistical uncertainties of the
values of the rotational constants were taken as the 1$\sigma$ uncertainties of
the corresponding fits based on Eq.~\eqref{eq:lineposition}. In cases where the
$1\sigma$ value was less than $0.0005$~cm$^{-1}$, the standard deviation of the rotational constants
determined from the spectra of the $v^+=13 \leftarrow v'=10$ band
($0.0005$~cm$^{-1}$) was taken instead. The statistical uncertainties of the
band origins and rotational constants obtained from the spectra of the $v^+=13
\leftarrow v'=10$ bands of the $^{24}\mathrm{Mg}_2$ and
$^{24}\mathrm{Mg}^{26}\mathrm{Mg}$ isotopomers for different $N'$ values were
taken as the standard deviation of the mean.

Term values of the $\mathrm{X}^+\ ^2\Sigma_u^+(v^+)$ vibrational levels
relative to the $\mathrm{X}\ ^1\Sigma_g^+(v''=0)$ ground state of the neutral
molecule were determined using the term values of the $\mathrm {A}\
^1\Sigma_u^+(v')$ levels calculated using the potential-energy function of
Ref.~\citenum{knockel14}. The mean precision of 0.025~cm$^{-1}$ stated by the
authors of \cite{knockel14} was taken as the uncertainty of these A-state term
values.

\begin{table}
	\centering
	\renewcommand{\arraystretch}{1.4}
	\caption{Band origins $\tilde{\nu}_{v^+v'}$ of the $\mathrm{X}^+\ ^2\Sigma_u^+(v^+) \leftarrow \mathrm{A}\ ^1\Sigma_u^+(v')$ ionizing transitions of $^{24}\mathrm{Mg}_2$, term values $T_{v^+}$ of the $\mathrm{X}^+\ ^2\Sigma_u^+(v^+)$ vibrational levels relative to the ground rovibronic state of $^{24}\text{Mg}_2$ ($\mathrm{X}\ ^1\Sigma_g^+(v''=0)$) and rotational constants $B^+_{v^+}$ of the $\mathrm{X}^+\ ^2\Sigma_u^+(v^+)$ levels of $^{24}\text{Mg}_2^+$. The numbers in parentheses represent one standard deviation in units of the last digit. Values in italics are for the $^{24}\text{Mg}^{26}\text
{Mg}$ isotopomer.}
	\label{tab:lines}
	\vspace{0.1cm}
	\begin{tabular}{rrlll}
	\hline\hline
	 $v'$ &  $v^+$ &  $\tilde{\nu}_{v^+v'}$ / cm$^{-1}$ &   $T_{v^+}$ / cm$^{-1}$ & $B^+_{v^+}$ / cm$^{-1}$ \\
	\hline
	  5 &    3 & 25078.58(10) & 52136.52(10) & 0.1513(15) \\
	  5 &    4 & 25284.96(10) & 52342.90(10) &   0.150(2) \\
	    &	   & \textit{25285.6(5)} & \textit{52325.0(5)} & -- \\ 
	  7 &    5 & 25137.53(10) & 52547.36(10) & 0.1496(15) \\
	  7 &    6 & 25339.64(10) & 52749.47(10) &  0.1483(9) \\
	 10 &    7 & 25028.56(10) & 52949.62(10) &  0.1461(5) \\
	 10 &    8 & 25226.46(10) & 53147.52(10) &  0.1453(5) \\
	 10 &    9 & 25422.21(10) & 53343.27(10) &  0.1440(5) \\
	 10 &   10 & 25615.86(10) & 53536.92(10) &  0.1427(5) \\
	 10 &   11 & 25807.46(10) & 53728.52(10) &  0.1414(9) \\
	 10 &   12 & 25996.97(10) & 53918.03(10) &  0.1409(5) \\
	 10 &   13 &  26184.22(6) &  54105.28(6) &  0.1387(5) \\
	    &      & \textit{26169.42(5)}  & \textit{54540.41(6)} & \textit{0.1358(11)} \\
	 10 &   14 & 26369.64(10) & 54290.70(10) &  0.1386(5) \\
	\hline\hline
	\end{tabular}
\end{table}

The absolute assignment of the vibrational quantum numbers of the X$^+
(v^+)$ levels was obtained in a standard analysis of the isotopic
shifts~\cite{herzberg50} from spectra of the $\mathrm{X}^+\ ^2\Sigma_u^+
(13) \leftarrow \mathrm{A}\ ^1\Sigma_u^+(10)$ and $\mathrm{X}^+\ ^2\Sigma_u^+
(4) \leftarrow \mathrm{A}\ ^1\Sigma_u^+(5)$ transitions recorded for the $^
{24}\text{Mg}_2$ and $^{24}\text{Mg}^{26}\text{Mg}$ isotopomers.

\subsection{Molecular constants}

Molecular constants characterizing the $\mathrm{X}^+\ ^2\Sigma_u^+$ ground
electronic state of Mg$_2^+$ were derived from the term values and rotational
constants obtained in this work (see Table~\ref{tab:lines}) and are listed in
Table~\ref{tab:molconstants}. The harmonic ($\omega_\mathrm{e}$) and anharmonic
($\omega_\mathrm{e}x_\mathrm{e}, \omega_\mathrm{e}y_\mathrm{e}$) vibrational constants of $^{24}\text{Mg}_2^+$ were
obtained in a least-squares fit based on the standard polynomial
expansion~\cite{herzberg50}

 \begin{align}
 	T_{v^+} =& T_\mathrm{e} + \omega_\mathrm{e}\left(v^+ +\frac12\right) - \omega_\mathrm{e}x_\mathrm{e}\left(v^+ +\frac12\right)^2  \nonumber\\ &+ \omega_\mathrm{e}y_\mathrm{e}\left(v^+ +\frac12\right)^3 .
 \end{align}
The adiabatic ionization energy $E_\mathrm{I}$ of $^{24}$Mg$_2$ ($\mathrm{X}\
^1\Sigma_g^+(v''=0)$) corresponds to $\frac{E_\mathrm{I}}{hc} = T_{v^+=0} = 51\,503.9(4)$~cm$^{-1}$. The equilibrium
rotational constant $B_\mathrm{e}$ and rotation-vibration coupling constant $\alpha_e$
were obtained from the rotational constants in a least-squares fit based on the
standard expression~\cite{herzberg50}

\begin{equation}
 	B^+_{v^+} = B_\mathrm{e} - \alpha_e \left( v^+ + \frac12\right) .
\end{equation}
Finally, the dissociation energy $D_0(\mathrm{X}^+)$ of the $\mathrm{X}^+\ ^2\Sigma_u^+$
state was determined to be $10\,572.3(6)$~cm$^{-1}$ from the thermodynamic cycle
\begin{equation}
	D_0(\mathrm{X}^+) = E_\mathrm{I}(\mathrm{Mg}) + D_0(\mathrm{X}) - E_\mathrm{I} ,
\end{equation}
using the dissociation energy of the $\mathrm{X}\ ^1\Sigma_g^+$ state of
$^{24}$Mg$_2$ reported by Kn\"ockel \textit {et al.}~\cite{knockel13}
($D_0(\mathrm{X})/(hc)=405.1(5)$~cm$^{-1}$) and the ionization energy of $^{24}$Mg ($E_\mathrm{I}(\mathrm{Mg})/(hc)= 61 671.05(3)$~cm$^
{-1}$~\cite{kramida20,li13}).

\begin{table}
	\centering
	\renewcommand{\arraystretch}{1.4}
	\caption{Comparison of the molecular constants of the $\text{X}^+\ ^2\Sigma^+_u$ ground electronic state of $^{24}\text{Mg}_2^+$ derived from the PFI-ZEKE-PE spectra with values obtained in \textit{ab initio} quantum chemical calculations~\cite{smialkowski20a,li13}. The molecular constants are in cm$^{-1}$ except $R_\mathrm{e}$ which is in Bohr radii ($a_0$). The numbers in parentheses represent one standard deviation in units of the last digit.}
	\label{tab:molconstants}
	\vspace{0.1cm}
 	\begin{tabular}{rS[table-format=6.8]S[table-format=6.3]S[table-format=6.4]}
 		\hline\hline
 		& {Present} & {Ref. \citenum{smialkowski20a}} & {Ref. \citenum{li13}} \\
 		\cmidrule(lr){2-2} \cmidrule(lr){3-3} \cmidrule(lr){4-4}
 		$E_\mathrm{I}/(hc)$          & 51503.9(4) & &\\
 		$D_0(\mathrm{X}^+)$ & 10572.3(6) & & \\
 		$D_\mathrm{e}(\mathrm{X}^+)$ & {10\,679.6(6)\,$^a$\phantom{.00}} & {10\,575\,$^b$\phantom{.}} & 10642\\
 		$\omega_\mathrm{e}$          & 215.25(14) & 215 & 215.49\\
 		$\omega_\mathrm{e}x_\mathrm{e}$       & 1.102(16) & & 0.59 \\
		$\omega_\mathrm{e}y_\mathrm{e}$       & 0.0015(6) & & \\
		$R_\mathrm{e}$               & 5.684(9) & 5.70 & 5.69 \\
		$B_\mathrm{e}$               & 0.1554(5) & 0.154 & 0.155 \\
		$\alpha_e$          & 0.00120(5) & & 0.0011 \\
 		\hline\hline
 	\end{tabular}

 	\begin{flushleft}
 		\footnotesize $^a$\ Estimated using $D_\mathrm{e} = D_0 + \omega_\mathrm{e}/2 - \omega_\mathrm{e}x_\mathrm{e}/4 + \omega_\mathrm{e}y_\mathrm{e}/8$\\
 		\footnotesize $^b$\ The potential-well depth of $10532$~cm$^{-1}$ reported in Ref.~\citenum{smialkowski20a} is a misprint~\cite{tomza21}.
 	\end{flushleft}
\end{table}

The present value of the dissociation energy is in agreement with, but 500
times more accurate than, the only other experimental value we could find in
the literature ($10\,200(300)$~cm$^{-1}$~\cite{ding93}). Molecular constants
from recent \textit{ab initio} calculations~\cite
{alharzali18,smialkowski20a,li13,ricca94} are consistent with the present
values (see Table~\ref{tab:molconstants} for a detailed comparison with the
results of Refs.~\citenum{li13,smialkowski20a}). Recent high-level \textit{ab
initio} calculations based on the procedure used to characterize the ground and
excited states of the Sr$_2^+$ ion~\cite{smialkowski20b} also yield values
($D_\mathrm{e} = 10667$~cm$^{-1}$ and $R_\mathrm{e} = 5.694\ a_0$~\cite{tomza21}) that are in
excellent agreement with our experimental values.

\section{Conclusions}

We reported an experimental study of the $\mathrm{X}^+\ ^2\Sigma_u^+$
electronic ground state of the Mg$_2^+$ ion. Neutral Mg$_2$ molecules in their
ground $\mathrm{X}\ ^1\Sigma_g^+(v''=0)$ vibronic state were produced in a
laser-ablation supersonic-expansion source and excited to high Rydberg states
below the $\mathrm{X}^+(v^+, N^+)$ ionization thresholds by resonant ($1+1'$)
two-photon absorption. The PFI-ZEKE-PE spectra of the $v^+=3-14$ vibrational
levels of the ground-state ion were recorded with rotational resolution.
Deviations of the line intensities from those predicted for direct
photoionization were observed and attributed to vibrational channel
interactions. Term values and rotational constants were extracted from the
spectra and a set of accurate molecular constants was derived. In particular,
the present work provides improved values of the adiabatic ionization energy of
$^{24}$Mg$_2$ ($51\,503.9(4)$~cm$^{-1}$) and of the ground-state dissociation
energy of $^{24}$Mg$_2^+$ ($10 572.3(6)$~cm$^{-1}$). The
experimental~\cite{ding93} and
theoretical~\cite{alharzali18,smialkowski20a,li13,ricca94,tomza21} data
available in the literature are consistent with the present
results. Finally, a band observed in the ($1+1'$) REMPI spectrum of the
$\mathrm{A}\ ^1\Sigma_u^+(v'=10)
\leftarrow \mathrm{X}\ ^1\Sigma_g^+ (v''=0)$ transition could not be assigned
to any multiphoton ionization process from ground-state Mg$_2$ molecules and
 was tentatively attributed instead to a transition from metastable molecules
 in the $(1)\ ^3\Sigma_u^+$ state.

The present work opens the possibility to study electronically excited states
of Mg$_2^+$, and in particular its Rydberg states, by isolated-core multiphoton
Rydberg dissociation~\cite{genevriez20}. The structure and dynamics of these
states are expected to differ from the Rydberg states of MgAr$^+$, which is one of the few molecular ions for which the Rydberg states are well characterized~\cite{wehrli20a,genevriez20a,wehrli21b,genevriez21c,wehrli21}. In
the case of MgAr$^+$, the ground state of the doubly charged ion (MgAr$^{2+}$)
to which the Rydberg series converge is thermodynamically
stable~\cite{wehrli21}. In contrast, the electronic ground state of Mg$_2^{2+}$
is unstable overall, and a local minimum in the Coulomb-repulsion-type
potential-energy function is predicted to support several metastable
vibrational states~\cite{hogreve04}. Rydberg series of Mg$_2^+$ converging to
these metastable Mg$_2^{2+}$ levels should exhibit potential-energy functions
and dynamical properties very different from those of MgAr$^+$, and their
systematic investigation is a perspective of future work.

\section{Acknowledgements}

We thank M.\,Tomza and D.\,Reich and C.\,Koch for providing us the data files
of the potential-energy functions they calculated for Mg$_2^+$ and Mg$_2$,
respectively. We also thank H.\,Kn\"ockel and E.\,Tiemann for assistance in
using the potential-energy functions they derived for the $\mathrm{X}$ and
$\mathrm{A}$ state of Mg$_2$. Finally, we thank Hansj\"urg Schmutz and
Josef\,A. Agner for their technical assistance. This work is supported
financially by the Swiss National Science Foundation (Grant No. 200020B--
200478) and the European Research Council through an ERC advanced grant (Grant
No. 743121) under the European Union's Horizon 2020 research and innovation
programme.

\end{document}